\documentclass[nofootinbib,aps]{revtex4}
\usepackage{amstext,amsmath,amssymb}

\newcommand{\Ref}[1]{(\ref{#1})}

\newcommand{\N}{\mathbb{N}}

\newcommand{\R}{\mathbb{R}}
\newcommand{\C}{\mathbb{C}}

\def\be{\begin{equation}}
\def\ee{\end{equation}}
\def\bes{\begin{eqnarray}}
\def\ees{\end{eqnarray}}

\def\w{\wedge}
\def\la{\langle}
\def\ra{\rangle}
\def\f{\frac}

\def\om{\omega}

\def\mm{{\cal M}}

\def\eps{\epsilon}

\newcommand{\lalg}[1]{\mathfrak{#1}}
\newcommand{\SU}{\mathrm{SU}}
\newcommand{\SL}{\mathrm{SL}}
\newcommand{\SO}{\mathrm{SO}}

\newcommand{\su}{\lalg{su}}
\renewcommand{\sl}{\lalg{sl}}

\newcommand{\so}{\lalg{so}}

\def\imm{\gamma}
\def\dd{\partial}
\def\G{{\cal G}}
\def\H{{\cal H}}
\def\hh{{\cal H}}
\def\aa{{\cal A}}
\def\ss{{\cal S}}
\def\ii{{\cal I}}
\def\aab{{\bf A}}
\def\slc{\SL(2,\C)}

\begin{document}

%%%%%%%%%%%%%%

\title{Towards a Covariant Loop Quantum Gravity}
\author{{\bf Etera R. Livine}\footnote{elivine@perimeterinstitute.ca}}
\affiliation{Perimeter Institute, 31 Caroline St N., Waterloo, ON, Canada N2L 2Y5}
\affiliation{Laboratoire de Physique, ENS Lyon, CNRS UMR 5672, 46 All\'ee d'Italie, 69364 Lyon Cedex 07}

%\author[E. R. Livine]{E. R. LIVINE\\
%Perimeter Institute (Canada) and Ecole Normale Sup\'erieure de Lyon (France)}
%\chapter{Towards a Covariant Loop Quantum Gravity} % chap ??
%\setcounter{page}{??}
%\setcounter{equation}{??}

\begin{abstract}
We review the canonical analysis of the Palatini action without going to the time gauge
as in the standard derivation of Loop Quantum Gravity. This allows to keep track of the
Lorentz gauge symmetry and leads to a theory of Covariant Loop Quantum Gravity. This new
formulation does not suffer from the Immirzi ambiguity, it has a continuous area spectrum
and uses spin networks for the Lorentz group. Finally, its dynamics can easily be related
to Barrett-Crane like spin foam models.
\end{abstract}

\maketitle

%%%%%%%%%%%%%%%%%%%%%%%%%%%%%%%%%%%%
% INTRODUCTION
\section{Introduction}
Over the recent years, Loop Quantum Gravity (LQG) has become a promising approach to
quantum gravity (see e.g. \cite{lqgthomas,lqgcarlo} for reviews). It has produced
concrete results such as a rigorous derivation of the kinematical Hilbert space with
discrete spectra for areas and volumes, the resulting finite isolated horizon entropy
counting and regularization of black hole singularities, a well-defined framework for a
(loop) quantum cosmology, and so on. Nevertheless, the model still has to face several
key issues: a well-defined dynamics with a semi-classical regime described by Newton
gravity law and general relativity, the existence of a physical semi-classical state
corresponding to an approximately flat space-time, a proof that the no-gravity limit of
LQG coupled to matter is standard quantum field theory, the Immirzi ambiguity,\dots Here,
we address a fundamental issue at the root of LQG, which is necessarily related to these
questions: {\it why the $\SU(2)$ gauge group of Loop Quantum Gravity?} Indeed, the
compactness of the $\SU(2)$ gauge group is directly responsible for the discrete spectra
of areas and volumes, and therefore is at the origin of most of the successes of LQG:
what happens if we drop this assumption?

As we will see, this leads to a theory of {\it Covariant Loop Quantum Gravity}
\cite{sergeicanonical,sergeiarea,sergeiquantum}, which uses the same techniques and tools
as LQG but whose gauge group is the Lorentz group $\SL(2,\C)$ instead of $\SU(2)$.

\medskip

Let us start by reviewing the general structure of LQG and how the $\SU(2)$ gauge group
arises. In a first order formalism, General Relativity (GR) is formulated in term of
tetrad $e$ which indicates the local Lorentz frame and a Lorentz connection $\om$ which
describes the parallel transport. The theory is invariant under local Lorentz
transformations and (space-time) diffeomorphisms.

The complex formulation of LQG is equivalent to that first order formalism. It is a
canonical formulation based on a splitting of the space-time as a spatial slice evolving
in time. The canonical variables are the Ashtekar variables: a self-dual complex
connection $A^{\rm Ash}$ and its conjugate triad field $E$. The theory is invariant under
the Lorentz group $\SL(2,\C)$ (seen as the complexified $\SU(2)$ group) and under
space-time diffeomorphisms. In these variables, GR truly looks like a $\SU(2)$ gauge
theory. The difficulty comes from reality constraints expressing that the imaginary part
of the triad field $E$ vanishes and that the real part of the connection $A^{\rm Ash}$ is
actually a function of the $E$. More precisely, on one hand, keeping the metric real
under the Hamiltonian flow requires that ${\rm Re}\,E\nabla[E,E]=0$ and, on the other
hand, the real part of $A^{\rm Ash}=\Gamma(E)+iK$ is the spin-connection $\Gamma(E)$
while its imaginary part is the extrinsic curvature. Such constraints must be taken into
account by the measure of the space of connection and render the quantization
complicated.

The real formulation of LQG came later as a way to avoid the reality constraint issue and
has now become the standard formulation of LQG. It uses the real Ashtekar-Barbero
connection $A^{\gamma}=\Gamma(E)+\gamma K$ and its conjugate triad field $E$. $\gamma$ is
called the {\it Immirzi parameter} and is an arbitrary real parameter. The theory is
derived from the original first order GR formulation in a particular (partial) gauge
fixing, the {\it time gauge}, which breaks the local Lorentz invariance down to a local
$\SU(2)$ gauge invariance. The theory then has a compact gauge invariance, is free from
complicated reality conditions and its Hamiltonian (constraint) can be regularized and
quantized. Nevertheless, it appears as the result of a gauge fixing. The natural question
is whether this affects the quantization or not: can we trust all the results of the real
LQG formulation? As we will see, considering $\SU(2)$ as the gauge group of GR instead of
the non-compact Lorentz group is related to several issues faced by the standard
formulation of LQG.

\begin{itemize}

\item Since we have chosen a particular gauge fixing, should not we take it into account
in the measure on the phase space through a Faddeev-Popov determinant? Would it not
change the spectrum of the observables of the theory? Moreover, does choosing the time
gauge constrain us to a specific class of measurements?

\item The Ashtekar-Barbero connection $A^{\gamma}$, on the spatial slice, is not the
pull-back of a space-time connection \cite{samuel}, since one can show that its holonomy
on the spatial slice depends on the embedding on that slice in space-time. This is true
unless the Immirzi parameter is taken equal to the purely imaginary values $\gamma=\pm i$
corresponding to the original self-dual Ashtekar connection. From that point of view, the
real connection $A^{\gamma}$ can not be considered as a genuine gauge field and $\SU(2)$
can not be viewed as the gauge group of gravity.

\item The Complex LQG formalism has a simple polynomial Hamiltonian constraint. On the other
hand, the real LQG formulation has an extra non-polynomial term. In fact, it seems we
trade the reality condition problem with the issue of a more complicated Hamiltonian.

\item There is a discrepancy with the standard spin foam models for GR. Spin foam models
have been introduced as discretization of the GR path integral seen as a constrained
topological theory \cite{SFreview}. They naturally appear as the space-time formalism
describing the evolution and dynamics of the LQG canonical theory. Nevertheless, they use
the Lorentz group as gauge group and therefore the quantum states of quantum geometry are
spin networks for the Lorentz group \cite{noncompact} instead of the standard $\SU(2)$
spin networks of LQG.

\item In three space-time dimension, the standard Loop Gravity quantization of
3d gravity has as gauge group the full Lorentz group and not only the little group of
spatial rotations. Indeed, in three space-time dimensions, the gauge group is always the
Lorentz group, $\SO(3)$ in the Riemannian version \cite{3d} and $\SO(2,1)$ in the
Lorentzian theory \cite{2+1d}. This allows a precise matching between the LQG framework
and the spin foam quantization for 3d gravity.

\item Finally, the real LQG formulation faces the issue of the Immirzi ambiguity:
$\gamma$ is an arbitrary unfixed parameter. It enters the spectrum of geometrical
observables such as areas and volumes (at the kinematical level). It is usually believed
that black hole entropy calculations should fix this ambiguity by requiring a precise
match with the semi-classical area-entropy law. More recently, $\gamma$ has been argued
to be related to parity violation when coupling fermions to gravity. Nevertheless, at the
level of pure gravity, there still lacks a clear understanding of the physical meaning of
$\gamma$: it does not change the classical phase space and canonical structure but leads
to unitarily inequivalent quantization (at the kinematical level). We can not forget the
possibility that this dependence on $\gamma$ might only be due to the choice of the time
gauge.

\end{itemize}

Here, we review a Lorentz covariant approach to Loop Quantum Gravity, which has been
coined {\it Covariant Loop Quantum Gravity}. It is based on an explicit canonical
analysis of the original Palatini action for GR without any time gauge first performed by
Alexandrov \cite{sergeicanonical}. The canonical variables are a Lorentz connection and
its conjugate triad (a 1-form valued in the Lorentz algebra). The states of quantum
geometry are Lorentz spin networks which reduce in a particular case to the standard
$\SU(2)$ spin networks.

The main difference with the standard LQG is a continuous spectrum for areas at the
kinematical level. The main advantages of the formalism is that the Immirzi ambiguity
disappears and it becomes possible to make contact between the canonical theory and spin
foam models. The main drawbacks of the approach are a non-compact gauge group and a
non-commutative connection. Finally, there is still a lot of work left in order to
precisely define the framework: rigourously define and study the Hilbert space (the
problem is to deal with the non-commutativity of the connection) and derive the dynamics
of the theory (quantize the Hamiltonian constraint and compare to the standard spin foam
models).

%%%%%%%%%%%%%%%%%%%%%%%%%%%%%%%%%%%%

%%%%%%%%%%%%%%
\section{Lorentz covariant Canonical Analysis}
%%%%%%%%%%%%%%%

%  \subsection{}
%  \paragraph{}

In a first order formalism, GR is formulated in term of the space-time connection
$\om=\om_\mu^{IJ} J_{IJ} dx^\mu$, defined as a $\so(3,1)$-valued 1-form, and the tetrad
field $e^I=e^I_\mu dx^\mu$. The space-time is a four-dimensional Lorentzian manifold
$\mm$ with signature $(-+++)$. $I,..$ are internal indices living in the tangent
Minkowski space. $\eta_{IJ}$ is the flat metric and $J_{IJ}$ are the Lorentz generators.
$\mu,..$ are space-time indices. The
{\it Palatini-Holst} action is
\cite{holst}:
\be
S[\om,e]=\int_{\cal M}\,\left[
\f{1}{2}\,\epsilon_{IJKL}e^I\w e^J \w F^{KL}(\om)
-\f{1}{\imm}\, e^I\w e^J \w F_{IJ}(\om)\right],
\ee
where $F(\om)=d\om+\om\w\om$ is the curvature tensor of the connection $\om$. The metric
is defined from the tetrad field as $g_{\mu\nu}=e^I_\mu e^J_\nu \eta_{IJ}$. The first
term of the previous action is the standard Palatini action. Its equations of motion are
equivalent to the usual Einstein equations when the tetrad is non-degenerate. The second
term actually has no effect on the equations of motion and thus does not matter at the
classical level. The coupling constant $\gamma$ is the Immirzi parameter.

The difficulty in the canonical analysis comes from the second class constraints. Indeed,
%neglecting the second term at first,
the canonically conjugate variable to the connection $\om_a^{IJ}$ is
$\pi^a_{IJ}=\eps^{abc}\eps_{IJKL}e^K_be^L_c$. These variables are unfortunately not
independent and they satisfy the {\it simplicity constraints}:
\be
\forall a,b,\, \eps^{IJKL}\pi^a_{IJ}\pi^b_{KL}=0.
\ee
These constraints are the non-trivial part of the canonical structure. Nevertheless,
Holst showed in \cite{holst} that, in the time gauge $e_{a0}=0$, the tetrad $e$ reduces
to a triad field $E$, the simplicity constraints do not appear and we recover the
canonical phase space and constraints of the real formulation of LQG with the
Ashtekar-Barbero connection $A^{(\gamma)}$ conjugated to $E$ and the Immirzi parameter
$\imm$.

The natural question is: how did the simplicity constraints go away? Barros pushed
Holst's analysis further and showed it is possible to solve these constraints explicitly
\cite{barros}. The phase space is parameterized by two couples of conjugate variables
$(A,E)$ and $(\chi,\zeta)$. The first couple of canonical variables is a generalization
of the Ashtekar-Barbero connection and triad. The new variable $\chi$ is the time normal
(or internal time direction) defined as the normalised space component (in the internal
indices) of the time component of the tetrad field: $\chi^i=-e^{0i}/e^{00}$. Finally, it
is possible to gauge fix the boost part of the Lorentz gauge symmetry by fixing $\chi=0$.
This is the {\it time gauge}. In this frame, we exactly retrieve the variables and
constraints of LQG. However, the price is the loss of the explicit Lorentz covariance of
the theory.

%\medskip

%%%
\subsection{Second class constraints and the Dirac bracket}
%%%

The strategy of {\it Covariant Loop Gravity} is to compute the whole set of second class
constraints, derive the associated Dirac bracket and then quantize the theory. Here, we
follow the canonical analysis of \cite{sergeicanonical}.

We start with a space-time ${\cal M}\sim\R\times \Sigma$ where we distinguish the time
direction from the three space dimensions. We decompose the tetrad field $e^I$
%(as a 1-form)
as:
\bes
e^0&=&N{\rm d}t+\chi_i E^i_a {\rm d}x^a  \nonumber \\
e^i&=&E^i_aN^a{\rm d}t+E^i_a{\rm d}x^a
\ees
where  $i=1..3$ is an internal index (space components of $I$) and $a$ is the space index
labelling the coordinates $x^a$. $N$ and $N^a$ are respectively the {\it lapse} and the
{\it shift}. $\chi^i$ indicates the deviation of the normal to the canonical hypersurface
from the time direction: the time normal is defined as the normalised time-like 4-vector
$\chi=(1,\chi_i)/\sqrt{1-|\vec{\chi}|^2}$.

Let's call $X,Y,..=1..6$ $\sl(2,\C)$-indices labelling antisymmetric couples $[IJ]$. We
define new connection/triad variables valued in
%the Lie algebra
$\sl(2,\C)$ instead of the standard $\su(2)$ of LQG. The Lorentz connection $A^X_a$ is:
\be
A^X=(\f{1}{2}\om^{0i},\f{1}{2}\epsilon^i_{jk}\om^{jk}).
\ee
Then we define a ``rotational" triad and a boost triad,
\be
R^a_X=(-\epsilon_i^{jk}E^i_a\chi_k,E^i_a),
\qquad
B^a_X=(\star R^a)_X=(E^i_a,\epsilon_i^{jk}E^i_a\chi_k),
\ee
where $\star$ is the Hodge operator on $sl(2,\C)$ switching the boost and rotation part
of the algebra. We further define the actual projectors on the boost and rotation sectors
of $\sl(2,\C)$, $(P_R)^X_Y=R_a^XR^a_Y,(P_B)^X_Y=B_a^XB^a_Y$:
%\be
$$
P_R=\left(
\begin{array}{cc}
-\frac{(\delta_a^b\chi^2-\chi_a\chi^b)}{1-\chi^2} &
-\frac{ {\eps_a}^{bc}\chi_c}{1-\chi^2} \\
-\frac{ {\eps_a}^{bc}\chi_c}{1-\chi^2} &
\frac{\delta_a^b-\chi_a\chi^b}{1-\chi^2}
\end{array} \right),
\quad P_B={\rm Id}-P_R, \quad P_BP_R=0.
$$
%(P_B)^X_Y=B_a^XB^a_Y=\left(
%\begin{array}{cc}
%\frac{\delta_a^b-\chi_a\chi^b}{1-\chi^2} &
%\frac{ {\eps_a}^{bc}\chi_c}{1-\chi^2} \\
%\frac{ {\eps_a}^{bc}\chi_c}{1-\chi^2} &
%-\frac{\delta_a^b\chi^2-\chi_a\chi^b}{1-\chi^2}
%\end{array} \right).
$P_R$ projects on  the subspace $\su(2)_\chi$ generating the rotations leaving the vector
$\chi$ invariant, while $P_B$ projects on the complementary subspace. The action then
reads:
\be
S=\int dtd^3x\,\left(
\left(B^a_X-\f{1}{\imm}R^a_X\right)\dd_tA^X_a
+\Lambda^X\G_X+{\cal N}^a\H_a+{\cal N}\H
\right).
\ee
The phase space is thus defined with the Poisson bracket,
\be
\left\{A^X_a(x),\left(B^b_Y-\f{1}{\imm}R^b_Y\right)(y)\right\}=
\delta^X_Y\delta^b_a\delta^{(3)}(x,y).
\ee
$\Lambda^X,{\cal N}^a,{\cal N}$ are Lagrange multipliers enforcing the first class
constraints:
\bes
\G_X&=&{\cal D}_A \left(B_X-\f{1}{\imm}R_X\right), \nonumber\\
\H_a&=& -\left(B^b_X-\f{1}{\imm}R^b_X\right)F_{ab}^X(A),\nonumber\\
\H&=&\f{1}{1+\frac{1}{\imm^2}}
\left(B-\f{1}{\imm}R\right)\left(B-\f{1}{\imm}R\right)F(A).
\ees
%The $\G$'s generate $\SL(2,\C)$ gauge transformations. The vector constraint $\H_a$
%imposes invariance under spatial diffeomorphisms (leaving the canonical hypersurface
%invariant). Finally, the scalar constraint $\H$ is called the Hamiltonian constraint
%generating the (time) evolution of the canonical variables.
%The reader can find explicit expressions in \cite{sergeicanonical,sergeiarea}
However, in contrast to the usual LQG framework, we also have {\it second class}
constraints:
\be
\phi^{ab}=(\star R^a)^XR^b_X=0, \qquad
\psi^{ab}\approx RR{\cal D}_AR.
\label{simple1}
\ee
The constraint $\phi=0$ is the simplicity constraint. The constraint $\psi=0$ comes from
the Poisson bracket $\{\H,\phi\}$ and is required in order that the constraint $\phi=0$
is preserved under gauge transformations (generated by $\G,\H_a,\H$) and in particular
under time evolution. $\psi$ corresponds to the reality constraint ${\rm
Re}\,E\nabla[E,E]=0$ of Complex LQG.

To solve the second class constraints, we define the {\it Dirac bracket}
$\{f,g\}_D=\{f,g\}-\{f,\varphi_r\}\Delta^{-1}_{rs}\{\varphi_s,g\}$ where the Dirac matrix
$\Delta_{rs}=\{\varphi_r,\varphi_s\}$ is made of the Poisson brackets of the constraints
$\varphi=(\phi,\psi)$.
%The Dirac matrix $\Delta_{rs}=\{\varphi_r,\varphi_s\}$ made of the Poisson brackets of
%the constraints $\varphi=(\phi,\psi)$ is of the following form
%\cite{sergeicanonical,sergeiarea}:
%$$
%\Delta=
%\mat{0}{D_1}{-D_1}{D_2},
%\quad
%\Delta^{-1}=\mat{D_1^{-1}D_2D_1^{-1}}{-D_1^{-1}}{D_1^{-1}}{0}.
%$$
%%which can be easily inversed,
%%$$
%%\Delta^{-1}_{rs}=\mat{D_1^{-1}D_2D_1^{-1}}{-D_1^{-1}}{D_1^{-1}}{0}.
%%$$
%The {\it Dirac bracket} is then defined as
%$\{f,g\}_D=\{f,g\}-\{f,\varphi_r\}\Delta^{-1}_{rs}\{\varphi_s,g\}$.
Following \cite{sergeicanonical,sergeiarea}, one then checks that the algebra of the
first class constraints is not modified. Defining smeared constraints, we find the
following Dirac brackets:
$$
{\cal G}(\Lambda)=\int_\Sigma\, \Lambda^X{\cal G}_X, \quad
\H(N )=\int_\Sigma\, N \H,
\quad
{\cal D}(\vec N)=\int_\Sigma\, N^a(\H_a+A_a^X{\cal G}_X),
$$
%\bes
%&&{\cal G}(\Lambda)=\int d^3x\, \Lambda^X{\cal G}_X, \qquad
%\H(N )=\int d^3x\, N \H,
%\nonumber \\
%&&{\cal D}(\vec N)=\int d^3x\, N^a(\H_a+A_a^X{\cal G}_X),
%\ees
$$
\begin{array}{ll}
\Bigl\{ {\cal G}(\Lambda_1) ,{\cal G}(\Lambda_2) \Bigr\}_D=
{\cal G}([\Lambda_1, \Lambda_2]),
&
\left\{ {\cal D}(\vec N) ,{\cal D}(\vec M) \right\}_D=
-{\cal D}([\vec N,\vec M ]), \\
\left\{ {\cal D}(\vec N) ,{\cal G}(\Lambda) \right\}_D=-
{\cal G}( N^a\partial_a\Lambda),
&
\Bigl\{ {\cal D}(\vec N) ,\H(N ) \Bigr\}_D=
-\H({\cal L}_{\vec N}N ),\\
\Bigl\{ \H(N ) ,{\cal G}(\Lambda) \Bigr\}_D =0,
&
\Bigl\{ \H(N ),\H(M ) \Bigr\}_D =
{\cal D}(\vec K)-{\cal G}(K^bA_b),
\end{array}
$$
$$
\begin{array}{ll}
[\Lambda_1, \Lambda_2]^X=f^X_{YZ}\Lambda_1^Y\Lambda_2^Z, &
[\vec N,\vec M ]^a= N^b\partial_bM^a-M^b\partial_bN^a,  \\
{\cal L}_{\vec N}N =N^a\partial_a N-N\partial_aN^a, &
K^b=(N\partial_aM-M\partial_aN)R^a_XR^b_Y g^{XY},
\end{array}
$$
where $f^X_{YZ}$ are the structure constant of the algebra $\sl(2,\C)$. With
$A\in\{1,2,3\}$ boost indices and $B\in\{4,5,6\}\sim\{1,2,3\}$ rotation indices, we have
$f^A_{AA}=f^A_{BB}=f^B_{AB}=0$ and $f^A_{AA}=-f^A_{AB}=-f^B_{BB}$ given by the
antisymmetric tensor $\eps$.
%$$
%\begin{array}{ccc}
%f_{A_1 A_2}^{A_3}=0,& f_{A_1 B_2}^{A_3}=-\eps^{A_1 B_2 A_3},&
%f_{B_1 B_2}^{A_3}=0, \\
%f_{B_1 B_2}^{B_3}=-\eps^{B_1 B_2 B_3},& f_{A_1 B_2}^{B_3}=0,& f_{A_1 A_2}^{B_3}=\eps^{A_1
%A_2 B_3}.
%\end{array}
%$$

The $\G$'s generate $\SL(2,\C)$ gauge transformations. The vector constraint $\H_a$
generates spatial diffeomorphisms on the canonical hypersurface invariant $\Sigma$.
Finally, the scalar constraint $\H$ is called the Hamiltonian constraint and generates
the (time) evolution of the canonical variables.

%%%
\subsection{The choice of connection and the Area spectrum}
%\subsection{Choosing the right connection}
%%%

As shown in \cite{sergeicanonical,sergeiarea,sergeiconnection}, although the triad field
$R$ is still commutative for the Dirac bracket, the properties of the connection $A$
change drastically: it is not canonically conjugated to the triad and it does not commute
with itself. Nevertheless, one should keep in mind that when using the Dirac bracket the
original canonical variables lose their preferred status and we should feel free to
identify better suited variables. Following \cite{sergeiconnection}, we do not modify the
triad $R$ but we look for a new connection ${\cal A}$ satisfying the following natural
criteria:
\begin{itemize}
\item[$\bullet$]
${\cal A}$ must be a Lorentz connection i.e. it should behave correctly under the Gauss
law $\G$:
\be
\{\G(\Lambda),{\cal A}^X_a\}_D=
\dd_a\Lambda^X-[\Lambda,A_a]^X
=\dd_a\Lambda^X-
f^X_{YZ}\Lambda^YA^Z_a.
\ee
\item[$\bullet$]
${\cal A}$ must be a 1-form et therefore properly transform under spatial
diffeomorphisms:
\be
\{{\cal D}(\vec{N}),{\cal A}\}_D=
{\cal A}^X_b\dd_aN^b-
N^b\dd_a{\cal A}^X_b.
\ee
\item[$\bullet$]
${\cal A}$ must be conjugated to the triad $R$. This is required in order that the area
operators $Area_{\cal S}\sim\int_{\cal S} d^2x\, \sqrt{n_an_bR^a_XR^{bX}}$ (with $n_a$
the normal to the surface ${\cal S}$) be diagonalized in the spin network basis resulting
from a loop quantization. This condition reads:
\be
\{{\cal A}^X_a(x),R^b_Y(y)\}_D\propto
\delta_a^b
\delta^{(3)}(x,y).
\ee
\end{itemize}
We obtain a 2-parameter family of such connections ${\cal A}(\lambda,\mu)$
\cite{sergeiconnection}:
\bes
{\cal A}_a^X(\lambda,\mu) &=&
A_a^X+\frac{1}{2}\left(\imm+\lambda - \mu \star\right)
P_R\f{\left(\imm-\star\right)}{1+\imm^2}\, [B_a,\G]^X
\nonumber \\
&&+ (\lambda + (1-\mu) \star)
\left( P_R\star A_a^X + \Theta_a^X(R)\right),
\label{connections}
\ees
with
$$
\Theta_a^X(R)=\Theta_a^X(\chi)=
\left(- \frac{\eps^{ijk}\chi_j\dd_a\chi_k}{1-\chi^2},
\frac{\dd_a\chi^i}{1-\chi^2} \right).
$$
Their commutation relation with the triad are very simple:
\be
\{ {\cal A}^X_a(\lambda,\mu),B_Y^b\}_D=\delta_a^b \left[\left(
\mu-\lambda\star \right) P_B\right]^X_Y
\ee
\be
\{ {\cal A}^X_a(\lambda,\mu),P_B\}_D=
\{ {\cal A}^X_a(\lambda,\mu),\chi\}_D=0.
\label{Achi}
\ee
Despite this, the bracket $\{{\cal A},{\cal A}\}_D$ remains complicated. From there, the
loop quantization chooses functions of ${\cal A}$ (Wilson loops and spin networks) as
wave functions and raises the triads $B,R$ to derivation operators. Each connection
${\cal A}(\lambda,\mu)$ will lead to a non-equivalent quantization. We can then compute
thethe action of an area operator on a ${\cal A}(\lambda,\mu)$ Wilson line and we find
\cite{sergeiconnection,sergeiquantum}:
%\be
$$
Area_{\cal S}\sim l_P^2 \sqrt{(\lambda^2 + \mu^2) C(\su(2)_\chi) -\mu^2 C_1(\sl(2,\C))
+\lambda\mu C_2(\sl(2,\C))}.
$$
%\ee
where $C(su(2)_\chi)=\vec{J}\cdot\vec{J}$ is the Casimir operator of $\su(2)_\chi$
(stabilizing the vector $\chi$), $C_1(sl(2,\C))=T^XT_X=\vec{J}^2-\vec{K}^2$ and
$C_2(sl(2,\C))=(\star T)^XT_Y=\vec{J}\cdot\vec{K}$  are the two (quadratic) Casimirs of
$\sl(2,\C)$. Since the algebra $\su(2)$ enters the formula, one could think at first that
this area spectrum is not Lorentz invariant. However, one must not forget that $\chi$
enters the formula and gets rotated under Lorentz transformations. Thus we see two
alternatives:
\begin{enumerate}
\item Either we work with functionals of the connection ${\cal A}$. Then a basis of
quantum states is provided by spin networks for the Lorentz group. These are labelled by
unitary representations of $\sl(2,\C)$, they diagonalize $C_1(\sl)$ and $C_2(\sl)$, but
they do not diagonalize $C(\su)$. Therefore they do not diagonalize the area operator.
\item Or we work with functionals of both the connection ${\cal A}$ and the time normal
field $\chi$. This is possible with ${\cal A}$ and $\chi$ commute (see \Ref{Achi}). It is
possible to introduce {\it projected spin networks}, which project on given eigenvalues
of $C(\su)$ and therefore diagonalize the area operator. We will discuss the details of
these states later.
\end{enumerate}
In the following, we will work with the latter alternative. Then the irreducible unitary
representations (of the principal series) of $\sl(2,\C)$ are labelled by a couple of
numbers $(n\in\N,\rho\ge0)$. The Casimir's values are then:
\be
C_1=n^2-\rho^2-1, \quad C_2=2n\rho, \quad C=j(j+1), \quad \textrm{with } j\ge n.
\ee
The restriction $j\ge n$ comes from the decomposition of the $\sl(2,\C)$ representations
on $\su(2)$ irreducible representations. Moreover this condition ensures that the area
eigenvalues are all real (and positive) for any value of $(\lambda,\mu)$. This is a nice
consistency check. Note however that, since the formula involves the real parameter
$\rho$, we lose the discreteness of the spectrum, which was a key result of LQG!

Now, it seems that we do not have any preferred choice of connection, and therefore no
rigorous prediction on the area spectrum. This would be an extra ambiguity besides the
choice of the Immirzi parameter $\gamma$. Instead, we choose to impose further
constraints on the connection ${\cal A}(\lambda,\mu)$ and two criteria naturally appear:
\begin{enumerate}
\item We require that the connection behaves properly under {\it space-time}
diffeomorphisms, generated by $\H_a$ and $\H$.
\item We require that the connection be commutative, i.e that $\{{\cal A},{\cal A}\}_D$
vanishes.
\end{enumerate}
Unfortunately, these two conditions are not compatible. As we will in the next sections,
the first choice corresponds to the only unique choice of a {\it covariant connection}
and is the one used by the proposed Covariant LQG. Very interestingly, the area spectrum
for this covariant connection does not depend on the Immirzi parameter $\gamma$. While
this resolves the Immirzi ambiguity, it is still complicated to quantize the theory due
to the non-commutativity of the connection. On the other hand, the second criteria leads
to a unique commutative Lorentz extension of the Ashtekar-Barbero connection. It allows
to recover the $\su(2)$ structure and area spectrum and Immirzi ambiguity of the real
formulation of LQG.

This raises the issue of the space-time covariance of the standard formulation of LQG
based on the Ashtekar-Barbero connection. Although there is no doubt that $\H_a$ and $\H$
satisfy the same algebra as the generators of the space-time diffeomorphisms, the action
of $\H$ on the connection is not the usual one. This means that this connection is not a
space-time 1-form and thus does not have a clear geometric interpretation. Although it is
not clear to which extent this is a problem, we expect this to be an obstacle when
studying the quantum dynamics of the theory.

%%%%%%%%%%%%%%%%%
\section{The Covariant Connection and Projected Spin Networks}
%%%%%%%%%%%%%%%%%

%%%
\subsection{A continuous area spectrum}
%%%

As shown in \cite{sergeiconnection,sergeiquantum}, there is a unique space-time
connection, i.e which transforms as a 1-form under space-time diffeomorphism generated by
the constraints $\H_a,\H$. It is actually the unique connection which is equal to the
original connection $A$ on the constrained surface $\G^X=\H_a=\H=0$.  It corresponds to
the choice $(\lambda,\mu)=(0,1)$ and we will simply write $\aa$ for $\aa(0,1)$ in the
following sections. Its brackets with the triad is:
\be
\{\aa^X_a,B^b_Y\}_D=\delta_a^b(P_B)^X_Y,\quad
\{\aa^X_a,(P_B)^Y_Z\}_D=0.
\ee
The first bracket says that only the boost part of the connection seems to matter. The
second relation is also very important and states that the field $\chi$ commutes with
both the connection and can thus be treated as an independent variable. Then, following
the results of the previous section, it turns out that the area spectrum does not depend
on the Immirzi parameter at all and is given by the boost part of the $\sl(2,\C)$
Casimir:
%\be
$$
{\rm Area}\sim l_P^2 \sqrt{C(\su(2)_\chi)-C_1(\sl(2,\C))}
\,=\,l_P^2\sqrt{j(j+1)-n^2+\rho^2+1}.
\label{airecov}
%\ee
$$
Interestingly, this spectrum is {\it not} the standard $\sqrt{j(j+1)}$ $\su(2)$-Casimir
area spectrum, but it contains a term coming from the Lorentz symmetry which makes it
{\it continuous}.

The problem with this connection is that it is non-commutative. Indeed, the bracket
$\{\aa^X,\aa^Y\}_D$ does not vanish and turns out to be complicated. At least, it is
possible to prove that it does not depend on the Immirzi parameter. Actually it was shown
\cite{sergeiquantum} that this complicated bracket was due to the fact that the
rotational part of $\aa$ was not independent from the triad field but equal to the
spin-connection:
$$
P_R\aa^X_a =\Gamma(R)^X_a
\sim [R,\dd R] + RR[R,\dd R].
%=\frac12 f^{W}_{YZ}(P_R)^{XY} R_a^Z \dd_c R^c_W + \frac12 f^{ZW}_Y\left(R^T_aR_{Tb}(P_R)^{XY}
%+R_b^XR_a^Y -R_a^XR_b^Y \right) R^c_Z \dd_c R^b_W.
$$
The explicit expression can be found in \cite{sergeiquantum,livinethese,sergeitime}. This
relation is reminiscent of the reality constraint of the complex LQG formulation where
the real part of the self-dual connection is a function of the triad $E$ and is
constrained to be the spin-connection $\Gamma(E)$. Moreover, it turns out that both the
rotation and the boost parts of the connection are commutative:
\be
\{(P_R\aa)^X,(P_R\aa)^Y\}_D=
\{(P_B\aa)^X,(P_B\aa)^Y\}_D=0.
\ee
At the end of the day, the non-commutativity of the connection comes from the facts that
$P_B\aa$ is canonically conjugate to the (boost) triad ($B=\star R$) and that the other
half of the connection $P_R\aa$ is a function of the triad. It thus seems as this
non-commutativity comes from taking into account the reality constraints.

%%%
\subsection{Projected spin networks}
%\subsection{Projected and simple spin networks}
%%%

In order to talk about the quantum theory and the area spectrum, we should precisely
define the Hilbert space and our quantum states of space(-time) geometry. Since geometric
observables (such as the area) involve $\chi$ and that $\chi$ commutes with $\aa$, it is
natural to consider functionals $f(\aa,\chi)$ as wave functions for the quantum geometry.
Then requiring gauge invariance under the Lorentz group $\SL(2,\C)$ reads:
\be
\forall g\in\slc\,f(\aa,\chi)=f({}^g\aa=g\aa g^{-1}+g\dd g^{-1},g.\chi).
\ee
Assuming that $\chi$ is time-like everywhere (i.e the canonical hypersurface is
space-like everywhere) and that all the fields are smooth, we can do a smooth gauge
transformation to fix $\chi$ to $\chi_0=(1,0,0,0)$ everywhere. Thus the wave function is
entirely defined by its section $f_{\chi_0}(\aa)=f(\aa,\chi_0)$ at $\chi=\chi_0$
constant:
$$
f(\aa,\chi)=f_{\chi_0}({}^g\aa) \quad \textrm{for all } g
\textrm{ such that } g.\chi=\chi_0.
$$
Then $f_{\chi_0}$ has a residual gauge invariance under $\SU(2)_{\chi_0}$. We are
actually considering functionals of the Lorentz connection $\aa$ which are not invariant
under the full Lorentz group $\SL(2,\C)$ but only under the {\it compact} group of
spatial rotations (defined through the field $\chi$).

To proceed to a loop quantization, we introduce {\it cylindrical functionals} which
depend on the fields $\aa,\chi$ through a finite number of variables. More precisely,
given a fixed oriented graph $\Gamma$ with $E$ links and $V$ vertices, a cylindrical
function depends on the holonomies $U_1,..,U_E\in\SL(2,\C)$ of $\aa$ along the edges of
$\Gamma$ and on the values $\chi_1,..,\chi_V$ of $\chi$ at the vertices of the graph. The
gauge invariance then reads:
\be
\forall\, k_v\in \slc, \,
%\forall \{k_v\}\in \slc^{\otimes V}, \,
\phi (U_e,..,\chi_v,..)=
\phi(k_{s(e)}U_ek_{t(e)}^{-1},..,
k_v.\chi_v,..),
\ee
where $s(e),t(e)$ denote the source and target vertices of an edge $e$. As previously,
such an invariant function is fully defined by its section $\phi_{\chi_0}(U_1,..,U_E)$ at
constant $\chi_1=..=\chi_V=\chi_0$. The resulting function $\phi_{\chi_0}$ is invariant
under $(\SU(2)_{\chi_0})^V$: we effectively reduced the gauge invariance from the
non-compact $\SL(2,\C)^V$ to the compact $(\SU(2)_{\chi_0})^V$.

Physically, the field $\chi$ describe the embedding of the hypersurface $\Sigma$ in the
space-time $\mm$. From the point of view of the cylindrical functionals, the embedding is
defined only at a finite number of points (the graph's vertices) and is left fuzzy
everywhere else. At these points, the normal to the hypersurface is fixed to the value
$\chi_v$ and the symmetry thus reduced from $\SL(2,\C)$ to $\SU(2)_{\chi_v}$.

Since the gauge symmetry is compact, we can use the Haar measure on $\SL(2,\C)$ to define
the scalar product on the space of wave functions:
\bes
\la\phi|\psi\ra &=&
\int_{\SL(2,\C)^E}\prod_e dg_e\,
\bar{\phi}(g_e,\chi_v)\psi(g_e,\chi_v) \nonumber\\
&=&\int_{\SL(2,\C)^E}\prod_e dg_e\,
\bar{\phi}_{\chi_0}(g_e)\psi_{\chi_0}(g_e).
\ees
The Hilbert space $H_\Gamma$ is finally defined as the space of $L^2$ cylindrical
functions with respect to this measure. A basis of this space is provided by the
{\it projected spin networks} \cite{livinethese,projected}. Following the standard
construction of spin networks, we choose one (irreducible unitary) $\SL(2,\C)$
representation $\ii_e=(n_e,\rho_e)$ for each edge $e\in\Gamma$. However, we also choose
one $\SU(2)$ representation $j_e^{(v)}$ for each link $e$ at each of its extremities $v$.
Moreover, we choose a $\SU(2)$ intertwiner $i_v$ for each vertex instead of a $\SL(2,\C)$
intertwiner. This reflects that the gauge invariance of the cylindrical function is
$\SU(2)^V$.

Let's call  $R^{(n,\rho)}$ the Hilbert space of the $\SL(2,\C)$ representation ${\cal
I}=(n,\rho)$ and $V^j$ the space of the $\SU(2)$ representation $j$. If we choose a
(time) normal $x
\in \SL(2,C)/\SU(2)$ and consider the subgroup $\SU(2)_x$ stabilizing $x$,  we can
decompose $R^{\cal I}$ onto the irreducible representations of $\SU(2)_x$:
\be
R^{(n,\rho)}=\bigoplus_{j\ge n} V^j_{(x)}.
\ee
Let's call $P_{(x)}^j$ the projector from $R^{(n,\rho)}$ onto $V^j_{(x)}$:
\be
P_{(x)}^j=\Delta_j\int_{SU(2)_x}dg\,\overline{\zeta}^j(g)D^{(n,\rho)}(g),
\ee
where $\Delta_j=(2j+1)$ is the dimension of $V^j$, the integration is over $\SU(2)_x$,
$D^{\ii}(g)$ is the matrix representing the group element $g$ acting on $R^{\ii}$ and
$\zeta^j$ the character of the $j$-representation. To construct a projected spin network,
we insert this projector at the end vertices of every link which allows to glue the
Lorentz holonomies to the $\SU(2)$ intertwiners. The resulting functional is:
\bes
\phi^{(\ii_e,j_e,i_v)}(U_e,\chi_v)&=&
\prod_v i_v\,\left[\bigotimes_{e\hookleftarrow v} |{\cal I}_e\chi_vj_e^{(v)}m_e\rangle \right]
\\
&&\prod_e \langle {\cal I}_e\chi_{s(e)}j_e^{(s(e))}m_e|D^{\ii_e}(U_e)|{\cal
I}_e\chi_{t(e)}j_e^{(t(e))}m_e\rangle, \nonumber
\ees
with an implicit sum over the $m$'s. $|{\cal I} xjm\rangle$ is the standard basis of
$V^j_{(x)}\hookrightarrow R^{\cal I}$ with $m$ running from $-j$ to $j$. In short,
compared to the usual spin networks, we trace over the subspaces $V^j_{(\chi)}$ instead
of the full spaces $R^{\cal I}$.

Using these projected spin networks allows to project the Lorentz structures on specific
fixed $\SU(2)$ representations. This allows to diagonalize the area operators.
Considering a surface $\ss$ intersecting the graph $\Gamma$ only on one edge $e$ at the
level of a (possibly bivalent) vertex, its area operator $\textrm{Area}_\ss$ will be
diagonalized by the projected spin network basis with the eigenvalues given above:
$$
\textrm{Area}_\ss\,|\phi_\Gamma^{(\ii_e,j_e,i_v)}\rangle \,=\,
l_P^2\sqrt{j_e(j_e+1)-n_e^2+\rho_e^2+1}\,|\phi_\Gamma^{(\ii_e,j_e,i_v)}\rangle.
$$

The procedure is now simple. Given a graph $\Gamma$ and a set of surfaces, in order to
have a spin network state diagonalizing the area operators associated to all these
surfaces, we simply need to project that spin network state at all the intersections of
the surfaces with $\Gamma$. If we want to obtain quantum geometry states diagonalizing
the areas of all the surfaces in the hypersurface $\Sigma$, we would need to consider a
``infinite refinement limit" where we project the spin network state at all points of its
graph $\Gamma$. Such a procedure is described in more details in
\cite{sergeiquantum,livinethese}. However, from the point of view that space-time is
fundamentally discrete at microscopic scales, it sounds reasonable to be satisfied with
quantum geometry states that diagonalize the areas of a discrete number of surfaces
(intersecting the graph at the points where we have projected the spin network states).
This is consistent with the picture that considering a projected spin network state, the
embedding of the hypersurface $\Sigma$ into the  space-time is only well-defined at the
vertices of the graph, where we know the time normal $\chi$: at all other points,
$\Sigma$ remains fuzzy and so must be the surfaces embedded in $\Sigma$.

%%%
\subsection{Simple spin networks}
%%%

Up to now, we have described quantum states $f(\aa)$ and the action of triad-based
operators on them. We should also define the action of connection-based operators. We
normally expect that $\aa$ would act by simple multiplication of a wave function
$f(\aa)$. Unfortunately, in our framework, $\aa$ does not commute with itself, so this
na\"\i ve prescription does not work. The point is that, due to the second class
constraint, the rotation part of the connection $P_R\aa$ is constrained and must be equal
to the spin connection $\Gamma[R]$ defined by the triad $R$. This reflects the reality
constraints of LQG. The natural way out is that we would like wave functions which do
{\it not} depend on $P_R\aa$ but only on $P_B\aa$. On such a state, the operator $\widehat{P_R\aa}$
will be defined as $\Gamma[\widehat{R}]$, while $\widehat{P_B\aa}$ will act simply by
multiplication. This is consistent with the Dirac bracket since $P_B\aa$ commute with
itself, $\{P_B\aa^X,P_B\aa^Y\}_D=0$.

One way to achieve this using projected spin network is to consider the case where we
project on the trivial $\SU(2)$ representation $j=0$. These are called {\it simple spin
networks}. To start with, a simple representation $\ii=(n,\rho)$ of the Lorentz group is
defined such that $C_2(\ii)=2n\rho$ vanishes: we only consider representations of the
type $(n,0)$ and $(0,\rho)$. Then for a representation $\ii$ to contain a
$\SU(2)$-invariant vector (corresponding to the $j=0$ sector), we must necessarily have
$n=0$. Therefore, simple spin networks use simple Lorentz representations of the
continuous type $\ii_{\rm simple}=(0,\rho)$. A simple spin network is defined by the
assignment of such representations $(0,\rho_e)$ to each edge $e$ of the graph $\Gamma$.
Since $\SU(2)$-intertwiners are trivial for the trivial representation $j=0$, the
functional then reads:
\be
\phi^{(\rho_e)}(U_e,\chi_v)\,=\,
\prod_e \la (0,\rho_e)\chi_{s(e)}j=0|U_e|(0,\rho_e)\chi_{t(e)}j=0 \ra.
\ee
Let's point out that in this special case of projected spin networks, we can consider
open graphs (with ``one-valent" vertices).

Simple spin networks are such that $\phi_\Gamma(U_e[\aa])$ does not depend on $P_R\aa$ at
the vertices $v$ of the graph $\Gamma$. In particular, considering two simple spin
networks $\phi,\phi'$ based on two graphs $\Gamma$ and $\Gamma'$ which only intersect at
mutual vertices, then $\phi_\Gamma(U_e[\aa])$ and $\phi'_{\Gamma'}(U_e[\aa])$ commute.

From there, we have two alternatives. Either we consistently project the spin network
states onto $j=0$ at every point of the graph so that they completely solve the second
class constraints. Or we can keep working with the present simple spin networks who only
solve the second class constraints at a discrete level. At the end of the day, it will be
these same simple spin networks which appear as kinematical geometry states in the spin
foam quantization as we will see later.

To summarize, we proved that the second class constraints are taken into account at the
quantum level by restricting the previous projected spin networks to be simple. The final
area spectrum taking into account the Lorentz gauge invariance and all the (kinematical)
constraints is purely continuous:
\be
\textrm{Area}_\ss\,|\phi_\Gamma^{(\rho_e)}\rangle \,=\,
l_P^2\sqrt{\rho_e^2+1}\,|\phi_\Gamma^{(\rho_e)}\rangle,
\ee
for a surface $\ss$ intersecting $\Gamma$ on the edge $e$.
%Let us nevertheless point out the shift $\rho^2\rightarrow\rho^2+1$ which leads to a
%non-vanishing minimal area $l_P^2$.
Nevertheless, the shift $\rho^2\rightarrow\rho^2+1$ still leads to a non-vanishing
minimal area $l_P^2$.

%%%%%%%%%%%%%%%%%
\section{Going down to $\SU(2)$ Loop Gravity}
%%%%%%%%%%%%%%%%%

As we have said earlier, there is a unique commutative Lorentz connection, which we will
denote $\aab$, and which corresponds to the choice $(\lambda,\mu)=(-\imm,0)$. It
satisfies the following commutation relations:
\be
\{\aab,\aab\}_D=0,
\qquad
\{\aab^X_a,R^b_Y\}_D\,=\,\imm\delta^b_a(P_R)^X_Y.
\ee
Intuitively, while $\aa$ was a pure boost connection, $\aab$ is a purely rotational
connection. More precisely, $\aab$ can be simply expressed in terms of the original
connection $A$ and the time normal field $\chi$:
\be
\left|
\begin{array}{ccl}
P_R\aab &=&P_R(1-\imm\star)A-\imm\Theta,\\
P_B\aab &=& \star\Theta(\chi)=\star(\chi\wedge\dd\chi)
\end{array}
\right.,
%\label{contraintesaab}
\ee
where $\Theta(\chi)$ was introduced earlier in eqn.(\ref{connections}). From this
expression, it is clear that $\aab$ is commutative and that $P_B\aab$ is not an
independent variable. Actually, in the time gauge where the field $\chi$ is taken
constant equal to $\chi_0$, $\aab$ reduces to the $\SU(2)$-connection of the real
Ashtekar-Barbero formalism. Then $\aab$ is the natural Lorentz extension of that
$\SU(2)$-connection \cite{sergeiquantum}. Finally, the area spectrum for this connection
reproduces exactly the standard spectrum:
\be
{\rm Area}_\ss \sim l_P^2\sqrt{C(\su(2)_\chi)}
\,=\,l_P^2\,\sqrt{j(j+1)}
\ee
In order to completely recover LQG, we still need to take care of the second class
constraints. To faithfully represent the Dirac bracket, we would indeed like wave
functions which do not depend on the boost part of the connection $P_B\aab$. We take this
into account in the scalar product. Instead of using the $\SL(2,\C)$-Haar measure, we can
restrict ourselves to the $\SU(2)$ subgroup. More precisely, we define the scalar product
using the $\chi=\chi_0$ section of the wave functions:
\be
\langle f | g\rangle=
\int_{[SU_{\chi_0}(2)]^E} dU_e \,\overline{f_{\chi_0}(U_e)}
g_{\chi_0}(U_e).
\label{su2prod}
\ee
We are considering the usual scalar product of LQG, and a basis is given by the standard
$\SU(2)$ spin networks. Nevertheless, it is now possible to go out of the time gauge and
describe the wave functions for arbitrary $\chi$ fields. More precisely, if
$f_{\chi_0}(A)$ is given by a $\SU(2)$ spin network, then $f(A,\chi)$ is a projected spin
network in the ``infinite refinement" limit. The interested reader will find more details
in \cite{sergeiquantum,livinethese}.

%%%%%%%%%%%%%%%%%
%\section{Projected Spin Networks}
%%%%%%%%%%%%%%%%%

%%%%%%%%%%%%%%%%%
\section{Spin Foams and the Barrett-Crane model}
%%%%%%%%%%%%%%%%%

Up to now, we have described the kinematical structure of Covariant Loop (Quantum)
Gravity. We still need to tackle the issue of defining the dynamics of the theory. On one
hand, one can try to regularize and quantize {\it \`a la} Thiemann the action of the
Hamiltonian constraint either on the covariant connection $\aa$ or the commutative
connection $\aab$. In this case, we will naturally have to study the volume operator and
face the usual ambiguities of LQG. On the other hand, one can turn to the {\it spin foam}
formalism. Spin foams have evolved independently but parallely to LQG. Inspired from
state sum models, they provide well-defined path integrals for ``almost topological"
theories, which include gravity-like theories. Moreover, they use the same algebraic and
combinatorial structures as LQG. In particular, spin networks naturally appear as the
kinematical states of the theory. From this perspective, spin foams allow a covariant
implementation of the LQG dynamics and a rigorous definition of the physical inner
product of the theory.

In three space-time dimensions, pure gravity is described by a BF theory and is purely
topological. The spin foam quantization is given by the Ponzano-Regge model \cite{PR}.
Its partition function defines the projector onto the gravity physical states i.e wave
functions on the moduli space of flat Lorentz connections.

In four space-time dimensions, it turns out that general relativity can be recast as a
{\it constrained BF theory}. One can quantize the topological BF theory as a spin foam
model and then impose the extra constraints directly on the partition function at the
quantum level (e.g \cite{spinfoam}). For 4d gravity, this leads to the Barrett-Crane
model \cite{BC}. There are of course ambiguities in the implementation of the
constraints, which lead to different versions of this model. We show below that the
Barrett-Crane model provide a dynamical framework for Covariant LQG.

%%%
\subsection{Gravity as a constrained topological theory}
%%%

Let us start with the Plebanski action:
\be
S[\omega,B,\phi]\,=\,\int_{\mathcal{M}}\left[
B^{IJ}\,\wedge\,F_{IJ}[\omega]\,-\frac{1}{2}\phi_{IJKL}\,B^{KL}\,\wedge\,B^{IJ}\right],
\ee
where $\om$ is the $\so(3,1)$ connection, $F[\om]=d_\om\om$ its curvature, $B$ a
$\so(3,1)$-valued 2-form and $\phi$ a Lagrange multiplier satisfying
$\phi_{IJKL}=-\phi_{JIKL}=-\phi_{IJLK}=\phi_{KLIJ}$ and $\phi_{IJKL}\epsilon^{IJKL}=0$.
The equations of motion are:
%\be
$$
d B\,+\,[\omega,B]=0, \quad
 F^{IJ}(\omega)\,=\,\phi^{IJKL}B_{KL},\quad
B^{IJ}\,\wedge\,B^{KL}\,=\,e\,\epsilon^{IJKL},
$$
%\ee
with $e=\frac{1}{4!}\epsilon_{IJKL}B^{IJ}\wedge B^{KL}$. When $e\ne0$, the constraint on
$B$ is equivalent to the simplicity constraint,
$\epsilon_{IJKL}B^{IJ}_{ab}B^{KL}_{cd}=\epsilon_{abcd}e$.
%    $\epsilon_{IJKL}B^{IJ}_{ab}B^{KLab}=0$

This constraint is satisfied if and only if there exists a real tetrad field
$e^{I}=e^{I}_{a}dx^{a}$ such that either $B=\pm e\w e$ (sector $I_\pm$) or
$B=\pm\star(e\w e)$ (sector $II_\pm$).
%such that one of the following equations is true:
%\bes &I_\pm&\;\;\;\;\;\;\;\;\;\;B^{IJ}\,=\,\pm\,e^{I}\,\wedge\,e^{J} \nonumber\\
%&II_\pm&\;\;\;\;\;\;\;\;\;\;B^{IJ}\,=\,\pm\,
%\frac{1}{2}\,\epsilon^{IJ}\,_{KL}e^{K}\,\wedge\,e^{L}
%.
%%\label{BFsectors}
%\ees
These four sectors are due to the symmetry of the constraints under $B\rightarrow (\star
B)$. The $\star$ operation allows to switch the sectors: $I_+\rightarrow II_+ \rightarrow
I_- \rightarrow II_- \rightarrow I_+$. Restricting ourselves to the $II_{+}$ sector, the
action reduces to $S=\int\star(e\w e)\wedge F$
%\be
%S\,=\,\int_{\mathcal{M}}\,\epsilon_{IJKL}\,e^{I}\,\wedge\,e^{J}\,\wedge\,F^{KL},
%\ee
and we recover general relativity in the first order formalism. A first remark is that we
still have to get rid of the $I_\pm$ and $II_-$ sectors in the path integral at the
quantum level. These are respectively related to the chirality of the 3-volume and to the
issue of time orientation \cite{qtetra,daniele}. A second remark is that taking a more
general constraint on the $\phi$ field, for instance
$a\phi_{IJ}\,^{IJ}+b\phi_{IJKL}\epsilon^{IJKL}=0$, we recover the Palatini-Holst action
for general relativity with Immirzi parameter
\cite{BFwithImm}.

%(e.g \cite{BFwithImm})

%Indeed, the solutions to the $B$-constraint become:
%$$
%B^{IJ}=\alpha *(e^I \w e^J) + \beta\, e^I \w e^J,
%\quad {\rm with} \quad
%\f{b}{a}=\f{\alpha^2-\beta^2}{4\alpha\beta}.
%$$

%%%
\subsection{Simple spin networks again}
%%%

The spin foam strategy is to first discrete and quantize the topological BF theory as a
state sum model, then to impose the $B$-constraints on the discretized partition
function.

In order to discretize the path integral, we choose a triangulation (or more generally a
cellular decomposition) of the 4d space-time gluing 4-simplices together. We then
associate the $B$ field to triangles, $B^{IJ}(t)=\int_t B^{IJ}$,
%$$
%B^{IJ}(t)=\int_t B^{IJ},
%$$
and the connection curvature to the dual surfaces. The simplicity constraint of the
2-form, $\epsilon_{IJKL}B^{IJ}_{ab}B^{KL}_{cd}=e\epsilon_{abcd}$, is then translated to
the discrete setting. For any two triangles $t,t'$, we have:
$$
\epsilon_{IJKL}\,B^{IJ}(t)\,B^{KL}(t')
%=V(t,t')
=\int_{t,t'}e d^2\sigma \w d^2\sigma'
=V(t,t'),
$$
%$V(t,t')=\int_{t,t'}e d^2\sigma \w d^2\sigma'$
where $V(t,t')$ is the 4-volume spanned by the two triangles. In particular, for any two
triangles who share an edge, we have:
\be
\epsilon_{IJKL}\,B^{IJ}(t)\,B^{KL}(t') =0.
\label{BCsimplicity}
\ee
These are the Barrett-Crane constraints which are implemented at the quantum level. More
preceisly, we associate a copy of the $\sl(2,\C)$-algebra to each triangle $t$ and we
quantize the $B^{IJ}(t)$'s as the Lorentz generators $J^{IJ}_t$. For a given triangle
$t$, the previous constraint for $t=t'$ becomes $\eps_{IJKL}J^{IJ}_t J^{KL}_t=0$, which
is the vanishing of the second Casimir $C_2(\sl(2,\C))=0$. This means that the
representation $\ii_t$ associated to a triangle $t$ must be {\it simple}: either
$(n_t,0)$ or $(0,\rho_t)$. The first Casimir $C_1=J_{IJ}J^{IJ}$ gives the (squared) area
of the triangle. For the discrete series, $C_1(n,0)=-n^2+1$ is negative and the triangle
is time-like. For the continuous series, $C_1(0,\rho)=\rho^2+1$ is positive and the
triangle is space-like. Thus we  recover the same simplicity of the $\sl(2,\C)$
representations as in Covariant LQG. The only difference is that we only consider
space-like triangles in the canonical framework, and therefore only obtain the $(0,\rho)$
representations. The time-like representations would naturally appear in the canonical
setting if considering a time-like normal $\chi$ (e.g \cite{sergeitime}). In the
following, we will restrict ourselves to the $(0,\rho)$ representations.

Coupling between different triangles happens at the level of tetrahedra: to each
tetrahedron is associated an intertwiner between the representations attached to its four
triangles. Solving the constraints $\eps_{IJKL}J^{IJ}_tJ^{KL}_{t'}=0$ for every couple of
triangles $(t,t')$ of the tetrahedron leads to a unique intertwiner. This Barrett-Crane
intertwiner $I_{BC}:\otimes_{t=1}^4R^{(0,\rho_t)}\rightarrow\C$ is the only
$\SU(2)$-invariant intertwiner:
\be
I_{BC}=\int_{\SL(2,\C)/\SU(2)}d\chi\,
\bigotimes_{t=1}^4 \,\left\la (0,\rho_t),\chi,j=0 \right|.
\ee
We recover the intertwiner structure of the simple spin networks introduced for Covariant
LQG. More precisely, the quantum geometry states associated to any space-like slice of
the triangulation in the Barrett-Crane model are simple spin networks
\cite{sergeiquantum,daniele}.

This makes the link between the kinematical states of the canonical theory and the spin
foam states. Then the transition amplitudes of the Barrett-Crane model can be translated
to the canonical context and considered as defining the dynamics of Covariant LQG.

%%%
\subsection{The issue of the second class constraints}
%%%

% $\{.,.\}_D$

In the previous spin foam quantization, we discretize and quantize the path integral for
general relativity. We have dealt with the simplicity constraint $B.(\star B)=0$ by
imposing on the path integral. A priori, this corresponds to the simplicity constraint
(\ref{simple1}), $\phi=R.(\star R)=0$ of the canonical analysis. However, it seems as we
are missing the other second class constraint $\psi\sim RR{\cal D}_A R$. The $\psi$
constraints are essential to the computation of the Dirac bracket: shouldn't we
discretize them too and include them in the spin foam model?

The spin foam point of view is that we have already taken them into account. Indeed, the
$\psi$ are secondary constraints, coming from the Poisson bracket ${\hh,\phi}$: at first,
$\phi=0$ is only imposed on the initial hypersurface and we need $\psi=0$ to ensure we
keep $\phi=0$ under the Hamiltonian evolution. On the other hand, the Barrett-Crane model
is fully covariant and  $\phi=0$ is directly imposed on all the space-time structures: we
have projected on $\phi=0$ at all stages of the evolution (i.e on all hypersurfaces). The
Barrett-Crane construction ensures that a simple spin network will remain a simple spin
network under evolution. In this sense, we do not need the secondary constraints $\psi$.
It would nevertheless be interesting to check that a discretized version of $\psi$
vanishes on the Barrett-Crane partition function.

%%%%%%%%%%%%%%%%%
%\section{And Quantum Causal Histories}
%%%%%%%%%%%%%%%%%

\section{Concluding Remarks}

Starting with the canonical analysis of the Palatini-Holst action, we have shown how the
second class constraints are taken into account by the Dirac bracket. Requiring a good
behavior of the Lorentz connection under Lorentz gauge transformations and space
diffeomorphisms, we obtain a two-parameter family of possible connection variables.
Requiring that the connection further behaves as a 1-form under space-time
diffeomorphisms, we obtain a unique covariant connection. This leads to a Covariant LQG
with ``simple spin networks" (for the Lorentz group), a continuous area spectrum and an
evolution dictated by the Barrett-Crane spin foam model. The theory turns out independent
from the Immirzi parameter. The main obstacle to a full quantization is that the
non-commutativity of this connection. This can be understood as reflecting the reality
conditions of the complex formulation of LQG. On the other hand, there exists a unique
commutative connection. It turns out to be a generalization of the Ashtekar-Barbero
connection of the real formulation of LQG. We further recover the $\SU(2)$ spin networks,
the standard discrete area spectrum and the usual Immirzi ambiguity.

It seems that Covariant LQG could help address some long-standing problems of the
standard formulation of LQG, such as the Immirzi ambiguity, the issue of the Lorentz
symmetry, the quantization of the Hamiltonian constraint and how to recover the
space-time diffeomorphisms at the quantum level.

Finally, a couple of issues which should be addressed within the Covariant LQG theory to
ground it more solidly are:
\begin{itemize}
\item a study of the 3-volume operator acting on simple spin networks.
\item a derivation of the spin foam amplitudes from the Covariant LQG Hamiltonian constraint,
possibly following the previous work in 3d gravity \cite{karim}.
\end{itemize}

%%%%%%%%%%%%%%%%%%%%%%%%%%%%%%%%%%%%%%%%%%%%%%%%%

%\section*{Appendix}

%\section*{Acknowledgements}

%bouhbouh

%%%%%%%%%%%%%%%%%%

%\clearpage
\end{document}